\documentclass[10pt,a4paper]{article}
\usepackage[cp1251]{inputenc}
\usepackage[english]{babel}
\usepackage{amssymb,graphicx}
\usepackage{amsmath,amsfonts}
\usepackage{fullpage}

\def\a{\alpha}

\def\G{\Gamma}
\def\d{\delta}

\def\e{\epsilon}

\def\la{\lambda}
\def\La{\Lambda}

\def\x{\xi}

\def\p{\pi}
\def\P{\Pi}

\def\r{\rho}

\def\t{\tau}

\def\pa{\partial}

\def\Ac{\mathcal{A}}
\def\Bc{\mathcal{B}}
\def\Cc{\mathcal{C}}

\def\Fc{\mathcal{F}}
\def\Oc{\hat{\mathcal{O}}}
\def\bLa{\bar{\Lambda}}
\def\l{\left}
\def\r{\right}
\def\sumlim{\sum\limits}
\def\prodlim{\prod\limits}

\def\ch{\mathop{\mathrm{ch}}\nolimits}

\renewcommand{\Im}{\mathop{\mathrm{Im}}\nolimits}

\def\be{\begin{equation}}
\def\ee{\end{equation}}
\def\bea{\begin{eqnarray}}
\def\eea{\end{eqnarray}}
\def\bmu{\begin{multline}}
\def\emu{\end{multline}}
\def\beg{\begin{gather}}
\def\eeg{\end{gather}}

\begin{document}
\title{Nekrasov prepotential with~fundamental matter from the~quantum spin chain}
\author{Yegor Zenkevich\thanks{E-mail: yegor.zenkevich@gmail.com} \medskip \\
{\small Moscow State University, Physical Department,}\\ \medskip
{\small{\it Vorobjevy Gory, Moscow, 119899, Russia}}\\
{\small Institute for Theoretical and Experimental Physics}\\ \medskip
{\small {\it Bolshaya Cheremushkinskaya 25, Moscow, 117218, Russia}}\\
{\small Institute for Nuclear Research of the Russian Academy of Sciences} \\
{\small {\it 60th October Anniversary Prospect 7a, Moscow, 117312, Russia}}}
\date{}
\maketitle
\vspace{-10.cm}
\begin{center}
\hfill ITEP/TH-2/11
\end{center}
\vspace{8.cm}
\begin{abstract}
Nekrasov functions were conjectured in~\cite{Mironov:2009uv} to be related to exact Bohr-Sommerfeld periods of quantum integrable systems. This statement was thoroughly checked for the case of the pure $SU(N_c)$ gauge theory in~\cite{Mironov:2009dv} and~\cite{Popolitov:2010bz}. Here we successfully perform a set of checks in the case of gauge group $SU(N_c)$ with additional $N_f$ fundamental hypermultiplets. We show that the Baxter equation for the spin chain gives the same quantum periods as the one for the Gaudin system in this case.
\end{abstract}
\section{Introduction}
Nekrasov functions are non-trivial generalizations of hypergeometric series originally introduced to regularize the integrals over instanton moduli space~\cite{Nekrasov:2003af}. Recently they appeared in various contexts in theoretical physics and mathematics most importantly in the famous AGT conjecture~\cite{Alday:2009aq}.

Here we are concerned with the conjecture~\cite{Nekrasov:2009rc} suggesting a connection between quantum integrable systems and Nekrasov prepotential. A concrete statement was formulated in~\cite{Mironov:2009uv}. It relates quantum deformed periods on the Seiberg-Witten spectral curve to the Nekrasov prepotential with only one \mbox{$\e$-parameter} turned on. The conjecture has been tested extensively for the pure $SU(N)$ gauge theory in~\cite{Mironov:2009dv} and~\cite{Popolitov:2010bz}. Here we consider this theory with additional $N_f$ matter hypermultiplets in fundamental representation. Some checks for $SU(2)$ theory with $N_f=4$ flavours were performed in~\cite{Maruyoshi:2010iu} and~\cite{Tai:2010ps} employing the Baxter equation for the Gaudin magnet as the quantum version of the spectral curve equation. Here we perform a thorough set of checks for various $N_c$ and $N_f$ using instead the Baxter equation for the XXX spin chain. We show that the two approaches give the same answers, at least for lower orders in~$\hbar$. Additionally we show that the conjecture holds to \emph{all} orders in $\hbar$ for the perturbative part of the prepotential in the case of $N_c = 2$ with up to four hypermultiplets. Let us note that the equivalence of the Gaudin systems and spin chains is not trivial even in the \emph{classical} limit, so it is remarkable that the two systems do give the same \emph{quantum} periods.

Our general claim is as follows. One starts with the equation for the Seiberg-Witten spectral curve, which in our case reads~\cite{Krichever:1996ut}
\be
K(p) - \frac{\bar{\La}}2 \l(K_+(p)e^{ix} + K_-(p)e^{-ix}\r)=0\,, \label{class}
\ee
where $\bar \La = \La^{N_c - \frac{N_f}{2}}$ and
\be
K(p)=u_N \prod\limits_{j=1}^{N_c} (p - \lambda_j)=\sum\limits_{j=1}^{N_c} u_j p^j\,, \notag
\ee
\be
\begin{array}{ccc}
K_+(p)= \prod\limits_{j=1}^{N_+} (p + m_j)\,,& K_-(p)= \prod\limits_{j=N_+ +1}^{N_f} (p + m_j)\,, & K_+(p) K_-(p) = \prod\limits_{j=1}^{N_f} (p + m_j) = \sum\limits_{j=1}^{N_f} v_j p^j\,. \notag
\end{array}
\ee
The distribution of masses between~``$+$'' and~``$-$'' polynomials does not affect the periods of $p$. The periods of the differential $pdx$ found using~(\ref{class}) define the Seiberg-Witten prepotential~$\Fc_{SW}(a)$~(which we call the ``classical'' one):
\bea
a_i &=& \P^0_{A_i}(\la,\,m,\,\La) = \oint\limits_{A_i} p\,dx\,, \label{classper1}\\
\frac{\pa \Fc_{SW}}{\pa a_i} &=& \P^0_{B_i}(\la,\,m,\,\La) = \oint\limits_{B_i} p\,dx\,. \label{classper2}
\eea
One obtains the ``quantum deformed'' prepotential by quantization of Eq.~(\ref{class}). This means that Eq.~(\ref{class}) is understood as the semiclassical limit of Schr\"odinger-like equation (which is in fact the Fourier transform of the Baxter equation for the XXX spin chain) and $p(x)$ is just the classical momentum~\cite{Mironov:2009ib}. The periods of $pdx$ are the same as the periods used in the Bohr-Sommerfeld quantization conditions in ordinary quantum mechanics. They are therefore subject to quantum corrections: the differential $pdx$ is replaced by $Pdx = \sum_n \hbar^n p_n dx$. These \emph{exact quantum periods} determine the deformed prepotential by the formulas similar to~(\ref{classper1}),~(\ref{classper2}):
\bea
a_i &=& \P^\hbar_{A_i}(\la,\,m,\,\La) = \oint\limits_{A_i} P\,dx\,, \\
\frac{\pa \Fc_{Nek}}{\pa a_i} &=& \P^\hbar_{B_i}(\la,\,m,\,\La) = \oint\limits_{B_i} P\,dx\,.
\eea
The prepotential defined in this way turns out to coincide with the Nekrasov function with only one $\e$-parameter turned on,~${\Fc_{Nek}(a|\e_1=\hbar,\,\e_2=0)}$.

When one tries to quantize Eq.~(\ref{class})~--- that is write $-i\hbar \pa / \pa x$ instead of $p$~--- one encounters the operator ordering problem. This is in fact a minor obstacle because different orderings are related by the shift in masses of the hypermultiplets:
\be
e^{itx} (-i\hbar \pa + m_f) e^{i(1-t)x} = (-i\hbar \pa + m_f - t\hbar) e^{ix}\,.
\ee
Having this in mind one writes out the Baxter equation corresponding to Eq.~(\ref{class}):
\be
\l(K(-i \hbar \pa) - \frac{\bar{\La}}{2} \l( e^{\frac{ix}{2}} K_{+}(-i \hbar \pa) e^{\frac{ix}{2}} + e^{-\frac{ix}{2}} K_{-}(-i \hbar \pa) e^{-\frac{ix}{2}} \r) \r) \exp \left(\frac{i}{\hbar} \int^x P(y)\,dy \right)=0\,. \label{equation}
\ee
Here $P(y)$ is understood as a series $P = \sum_n (-i\hbar)^n p_n$, and $p_0 = p$.
We used our freedom to choose the ordering in Eq.~(\ref{equation}) (or equivalently to choose what we call the mass of the hypermultiplet) so as to make all the odd powers of $\hbar$ in periods vanish. This was automatic in the pure $SU(N)$ case but requires fixing the particular (symmetric) ordering in our case.

Let us briefly recall the idea of perturbative checks analogous to~\cite{Mironov:2009dv}. The algorithm is as follows:
\begin{itemize}
\item Solve the quantum Baxter equation perturbatively in $\hbar$ to find several first $p_n$. They will depend on the parameters $\la_i$, $m_f$, $\La$.
\item Rewrite $p_n$ as a differential operator in $\la_i$, $m_f$, $\La$ acting on $p_0$:
\be
p_n = \Oc_n p_0
\ee
\be
P = \sumlim_n (-i\hbar)^n \Oc_n p_0 = \Oc p_0
\ee
\item Act on the known classical periods $\P^0_{A_i}(\la,\,m,\,\La)$ with the operator $\Oc$ to obtain $\P^\hbar_{A_i}(\la,\,m,\,\La)$ and substitute it into the known derivatives of the Nekrasov prepotential $\left.\frac{\pa \Fc_{Nek}}{\pa a_i}\right|_{a_i= \P^\hbar_{A_i}(\la,\,m,\,\La)}$.
\item Act on the known classical periods $\P^0_{B_i}(\la,\,m,\,\La)= \left.\frac{\pa \Fc_{SW}}{\pa a_i}\right|_{a_i= \P^0_{A_i}(\la,\,m,\,\La)}$ with $\Oc$ to obtain $\P^\hbar_{B_i}(\la,\,m,\,\La)$.
\item Compare $\left.\frac{\pa \Fc_{Nek}}{\pa a_i}\right|_{a_i= \Oc \P^0_{A_i}(\la,\,m,\,\La)}$ with $\Oc \left.\frac{\pa \Fc_{SW}}{\pa a_i}\right|_{a_i= \P^0_{A_i}(\la,\,m,\,\La)}$.
\end{itemize}
In Section~\ref{baxter} we solve the quantum Baxter equation and find the differential operator $\Oc$ perturbatively for the first orders in~$\hbar$. In Section~\ref{Nek} we remind the Nekrasov functions relevant for our case. In Section~\ref{pertchecks} we perform the checks for lower orders in $\hbar$.

This prescription is of use to check the conjecture to any desired order both in $\hbar$ and $\La$, but the \emph{general} order cannot be approached in such a manner. However there exists a way to check the conjecture for 1-loop part of the Nekrasov function to \emph{all} orders in $\hbar$~\cite{Marshakov:2010fx}.

The 1-loop part of $\Fc_{Nek}$ is dominant for small $\La$, when all the instanton corrections are suppressed. Our objective is therefore to investigate the corresponding case $\bar{\La} \to 0$ in the quantum Baxter equation~(\ref{equation}). There are no instanton corrections to $\P^0_{A_i}$, so $\P^0_{A_i} = \la_i$ in this regime. Moreover, $\P^\hbar_{A_i}$ does not recieve $\hbar$ corrections and therefore $\P^\hbar_{A_i} = \P^0_{A_i} = \la_i$. This can be inferred from the following argument. In gauge theory the Seiberg-Wittem prepotential is defined up to addition of a function quadratic in $a_i$, thus to the classical B-periods one can add a term linear in $a_i$ (in zero instanton approximation just $\la_i$). As we have mentioned, periods of the Nekrasov prepotential can be generated by acting on classical periods with certain differential operator $\Oc$. Higher orders of $\Oc \P^0_{B_i}$ in $\hbar$ should not change if we consider $\P^0_{B_i} + \xi^j_i \la_j$ instead of $\P^0_{B_i}$ as a classical period. Thus the differential $\Oc$ does not change $\la_i$ and hence the A-periods in the zero instanton approximation.

To find the B-periods one observes that $P(x)$ can be rewritten as
\be
P(x) = -i\hbar \, \pa \ln \psi(x)\,,
\ee
where
\be
\psi(x) = \exp \left(\frac{i}{\hbar} \int^x P(y)\,dy \right) \,.
\ee
The B-periods are thus rewritten using $\psi(x)$:
\be
\P^\hbar_{B_i}(\la,\,m,\,\La) = \ln \psi(a) - \ln \psi(b)\,,
\ee
where $a$ and $b$ lie on different sides of the cut\footnote{Somewhat similar approach was used in~\cite{He:2010zzc} to calculate first orders in $\hbar$ expansion of the periods. Here we provide an exact result for~$\bar{\La} \to 0$.}. Solving the Baxter equation~(\ref{equation}) in the limit $\bar{\La} \to 0$ one finds agreement with perturbative part of the Nekrasov prepotential. We carry out this checks in the case of pure $SU(2)$ gauge theory with up to four hypermultiplets in Section~\ref{exactcheck}. The conclusions are presented in Section~\ref{conclusions}.

\section{Solving the Baxter equation}
\label{baxter}
Let us remind that the exact momentum $P$ is understood as a perturbative series $P=\sum p_n (-i \hbar)^n$. Therefore, one needs to rewrite Eq.~(\ref{equation}) in terms of $p_n$.
For this task it is convenient to define the generating function
\begin{multline}
G(\a) = \sum\limits_{n \geq 0} \exp \left(-\frac{i}{\hbar} \int^x P(y)\,dy \right) (-i\hbar \pa)^n \exp \left(\frac{i}{\hbar} \int^x P(y)\,dy \right) \frac{\a^n}{n!}=\\
=\exp \left(-\frac{i}{\hbar} \int^x P(y)\,dy \right) \exp \left(-i\hbar \a \pa \right) \exp \left(\frac{i}{\hbar} \int^x P(y)\,dy \right)= \exp \left(-\frac{i}{\hbar} \int^{x- i \a \hbar}_x P(y)\,dy \right)=\\
=\exp \left( \sum\limits_{k \geq 0} \frac{(-i\hbar)^k \a^{k+1}}{(k+1)!} \pa^k P(x) \right)\,. \label{genfunc}
\end{multline}
Substituting the expansion for $P$ into Eq.~(\ref{genfunc}) and expanding the exponent in terms of $(-i\hbar)$ one obtains
\begin{multline}
G(\a) = \exp \left( \sum\limits_{k \geq 0} \sum\limits_{i \geq 0} \frac{(-i\hbar)^{k+i} \a^{k+1}}{(k+1)!} \pa^k p_i(x) \right) = \\
= e^{\a p_0}  \sum\limits_{N \geq 0} (-i \hbar)^N  \sum\limits_{\substack{ n \geq 0 \\ l_j \geq 1 \\ l_1+ \ldots +l_n =N }} \sum\limits_{\substack{ 0 \geq i_j \geq l_j \\ i_1+ \ldots +i_n = M }} \a^{N-M+n} \frac{1}{n!} \prod\limits_{j=1}^n \frac{\pa^{l_j-i_j} p_{i_j}}{(l_j-i_j+1)!}=\\
=\sum\limits_{N \geq 0} (-i \hbar)^N  \sum\limits_{\substack{ n \geq 0 \\ l_j \geq 1 \\ l_1+ \ldots +l_n =N }} \sum\limits_{\substack{ 0 \geq i_j \geq l_j \\ i_1+ \ldots +i_n = M }}  \frac{1}{n!} \prod\limits_{j=1}^n \frac{\pa^{l_j-i_j} p_{i_j}}{(l_j-i_j+1)!} \l( \frac{d}{dp_0} \r)^{N-M+n} e^{\a p_0} \,.
\end{multline}
Expanding $G(\a)$ in terms of $\a$ one gets
\begin{multline}
\exp \left(-\frac{i}{\hbar} \int^x P(y)\,dy \right) (-i\hbar \pa)^L \exp \left(\frac{i}{\hbar} \int^x P(y)\,dy \right) = \\
= \sum\limits_{N \geq 0} (-i \hbar)^N  \sum\limits_{\substack{ n \geq 0 \\ l_j \geq 1 \\ l_1+ \ldots +l_n =N }} \sum\limits_{\substack{ 0 \geq i_j \geq l_j \\ i_1+ \ldots +i_n = M }}  \frac{1}{n!} \prod\limits_{j=1}^n \frac{\pa^{l_j-i_j} p_{i_j}}{(l_j-i_j+1)!} \l( \frac{d}{dp_0} \r)^{N-M+n}  p_0^L
\end{multline}
and finally
\begin{multline}
\exp \left(-\frac{i}{\hbar} \int^x P(y)\,dy \right) K(-i\hbar \pa) \exp \left(\frac{i}{\hbar} \int^x P(y)\,dy \right) = \\
= \sum\limits_{N \geq 0} (-i \hbar)^N  \sum\limits_{\substack{ n \geq 0 \\ l_j \geq 1 \\ l_1+ \ldots +l_n =N }} \sum\limits_{\substack{ 0 \geq i_j \geq l_j \\ i_1+ \ldots +i_n = M }}  \frac{1}{n!} \prod\limits_{j=1}^n \frac{\pa^{l_j-i_j} p_{i_j}}{(l_j-i_j+1)!} \l( \frac{d}{dp_0} \r)^{N-M+n} K(p_0)\,. \label{K(d)}
\end{multline}
Now having the expression~(\ref{K(d)}) we can proceed to solve Eq.~(\ref{equation}).
One notices that the system of equations is ``triangular'' (though nonlinear)~--- meaning that $p_N$ appears linearly in order $(-i\hbar)^N$ and is expressed through the lower $p_m$. Thus for any given $N$ one can solve the equation iteratively.

The first correction $p_1$ turns out to be total derivative:
\be
p_1=-\frac12 \frac{\pa}{\pa x} \ln (K' + V')\,,
\ee
where $V=V(x,p_0(x))=-\frac{\bar{\La}}{2}(K_+ e^{ix} +K_- e^{-ix})$, $V'=\frac{\pa V}{\pa p_0}$ and $K'=\frac{\pa K}{\pa p_0}$. As explained in the introduction, this was exactly our aim when we chose operator ordering on Eq.~(\ref{equation}). The non-trivial correction appears only in the second order:
\begin{multline}
p_2 =
-\frac{\pa V' \pa V''}{12 (K'+V')^2}+\frac{\pa V'^2 (K''+V'')}{24 (K'+V')^3}+\frac{(K''+V'')^2 V}{24 (K'+V')^3}-\frac{(K'''+V''') V}{24 (K'+V')^2}-\frac{(K''+V'') V'}{24 (K'+V')^2}\,.
\end{multline}
The third order is again total derivative as required, though less simple:
\begin{multline}
p_3= \frac{\pa}{\pa x} \Biggl( -\frac{\pa V' \pa V'''}{24 (K'+V')^2}-\frac{\pa V \pa V''''}{24 (K'+V')^2}+\frac{\pa V'
\pa V'' (K''+V'')}{8 (K'+V')^3}+\\
\frac{\pa V \pa V''' (K''+V'')}{6 (K'+V')^3}-\frac{3 \pa V'^2 (K''+V'')^2}{16 (K'+V')^4}-\frac{3
\pa V \pa V'' (K''+V'')^2}{8 (K'+V')^4}+\\
\frac{5 \pa V \pa V' (K''+V'')^3}{8 (K'+V')^5}-\frac{5 \pa V^2 (K''+V'')^4}{16 (K'+V')^6}+\frac{5
\pa V'^2 (K'''+V''')}{48 (K'+V')^3}+\\
\frac{\pa V \pa V'' (K'''+V''')}{6 (K'+V')^3}-\frac{2 \pa V \pa V' (K''+V'') (K'''+V''')}{3
(K'+V')^4}+\\
\frac{25 \pa V^2 (K''+V'')^2 (K'''+V''')}{48 (K'+V')^5}-\frac{\pa V^2 (K'''+V''')^2}{12 (K'+V')^4}+\\
\frac{\pa V \pa V' (K^{(4)}+V^{(4)})}{8 (K'+V')^3}-\frac{7 \pa V^2 (K''+V'') (K^{(4)}+V^{(4)})}{48
(K'+V')^4}+\frac{\pa V^2 (K^{(5)}+V^{(5)})}{48 (K'+V')^3}+\\
\frac{(K''+V'')^3 \pa V}{8 (K'+V')^4}-\frac{(K''+V'') (K'''+V''') \pa V}{6 (K'+V')^3}+\frac{(K^{(4)}+V^{(4)})
\pa V}{24 (K'+V')^2}-\\
\frac{(K''+V'')^2 V'}{8 (K'+V')^3}+\frac{(K'''+V''') V'}{12 (K'+V')^2}+\frac{(K''+V'')
V''}{16 (K'+V')^2}-\frac{V'''}{48 (K'+V')} \Biggr)\,.
\end{multline}
This conforms with our intention of eliminating all the odd orders of $\hbar$ in the quantum periods.
\subsection{The differential operator}
\label{diffop}
It is instructive to see that the classical periods $\P^0$ do not depend on the particular partition of $m_j$ into the ``$+$'' and ``$-$'' groups. Moving $m_k$ from one group to another is equivalent to shifting $x \mapsto x - i\ln (p_0 + m_k)$. Then the change in $\P^0$ is
\be
\P^0 = \oint p_0 \, dx \mapsto  \oint p_0 \, dx - i \oint \frac{p_0 \,dp_0}{p_0 + m_k} = \P_0 - i \oint \frac{p_0 \,dp_0}{p_0 + m_k}\,.
\ee
Because the contours can be chosen not to encircle the $-m_k$ point, the period is left unchanged. If analogously to~\cite{Popolitov:2010bz} one considers a more general form of Eq.~(\ref{equation}) with two different constants (instead of single one~--- $\bLa$) in front of the two exponents then the classical periods $\P_0$ will depend only on the product of these constants.

Having these two facts in mind one can consider only the case $K_-=1$ and thus replace $\pa V^{(j)}$ by $i V^{(j)}$ for $j \geq 1$. We introduce differential operators
\bea
D_i &=& \sum\limits_{j=i}^{N_c} \frac{j!}{(j-i)!} u_j \frac{\pa}{\pa u_{j-i}}\,, \\
\tilde D_i &=& \sum\limits_{j=i}^{N_c} \frac{j!}{(j-i)!} v_j \frac{\pa}{\pa v_{j-i}} \quad \text{for} \quad i \geq 1\,,\\
\tilde D_0 &=& \bar{\La} \frac{\pa}{\pa \bar{\La}}\,.
\eea
It can be checked then that
\be
p_2 = - \frac{1}{24}(D_2 \tilde D_0 + \tilde D_2 \tilde D_0 - \tilde D_1 \tilde D_1) p_0\,. \label{p2}
\ee
\subsection{A-periods to the order $\La^{2N_c-N_f}$}
\label{Aper}
The classical $A$ periods were calculated in~\cite{D'Hoker:1996nv}
\begin{multline}
\P^0_{A_k} \equiv a_k = \la_k + \sum\limits_{m \geq 1} \frac{\bar \La^{2m}}{2^{2m} (m!)^2} {\l( \frac{\pa}{\pa \la_k} \r)}^{2m-1} \l( \frac{\prod\limits_{f=1}^{N_f} (\la_k + m_f)^m}{\prod\limits_{i \neq k}^{N_c} \la_{ki}^{2m}} \r)=\\
=\la_k + \frac{\bar \La^2}{4} \frac{\prod\limits_{f=1}^{N_f} (\la_k + m_f)}{\prod\limits_{i \neq k}^{N_c} \la_{ki}} \l( \sum\limits_{g=1}^{N_f} \frac{1}{(\la_k + m_g)} - 2 \sum\limits_{l \neq k}^{N_c} \frac{1}{\la_{kl}} \r) + O(\bar \La^4)\,. \label{Aperiods}
\end{multline}
\section{Nekrasov functions}
\label{Nek}
The Nekrasov function for the $SU(N_c)$ theory with $N_f$ fundamentals is given by~\cite{Alday:2009aq}\footnote{Notice that one must use the shifted masses $m_f-\frac{\e_1}{2}$.}
\be
\Fc_{Nek} = \Fc_{pert}^{Nek} + \Fc_{inst}^{Nek}\,.
\ee
The perturbative periods are
\begin{multline}
-\frac14 \frac{\pa \Fc_{pert}^{Nek}}{\pa a_i} = - (N_c - \frac{N_f}{2}) a_i \ln \La + \frac{\e_1}{2} \sum\limits_{j \neq i} \ln \l( \frac{\G \l(1+\frac{a_{ij}}{\e_1}\r)}{\G\l(1-\frac{a_{ij}}{\e_1}\r)}\r) - \frac{\e_1}{4} \sum\limits_{n} \ln \l( \frac{\G \l(1+\frac{a_i+m_n-\e_1/2}{\e}\r)}{\G\l(1-\frac{a_i+m_n-\e_1/2}{\e_1}\r)}\r)=\\
=\sum\limits_{j \neq i} a_{ij} \l( \ln \frac{a_{ij}}{\La} -1 + \sum\limits_{m \geq 1} \frac{B_{2m}}{2m(2m-1)} \l( \frac{\e_1}{a_{ij}} \r)^{2m} \r) - \\
-\frac12 \sum\limits_{n} (a_i+ m_n) \l( \ln \frac{a_i + m_n}{\La} - 1 + \sum\limits_{m \geq 1} \frac{B_{2m} (2^{-2m+1}-1)}{2m(2m-1)} \l( \frac{\e_1}{a_i+m_n} \r)^{2m} \r)\,. \label{Nekpert}
\end{multline}
For our purposes we need only 1-instanton part of the Nekrasov prepotential which is given by
\be
\Fc_{1-inst}^{Nek} = \frac{\La^{2N_c-N_f}}2 \sumlim_{i=1}^{N_c} \frac{\prodlim_{f=1}^{N_f} (\la_i+m_f+\frac{\e_1}2)}{\prodlim_{j \neq i} \la_{ij}(\la_{ij}+\e_1)}\,. \label{1instNek}
\ee
\section{Checks up to $\hbar^2$}
\label{pertchecks}
\subsection{Zero instanton approximation}
As was already mentioned in the introduction (and can be seen from the form of the operator~(\ref{p2})), A-periods do not receive any corrections in zero instanton approximation and thus
\be
\P^\hbar_{A_k} = \P^0_{A_k} = \la_k\,.
\ee
What we are going to check is that
\be
\hat O \P^0_{B_k} (\P^0_{A}) \stackrel{?}{=} \P^\hbar_{B_k} (\hat O \P^0_{A})\,,
\ee
where
\be
\P^0_{B_k} = - \frac14 \frac{\pa \Fc_{SW}}{\pa a_k}\,, \qquad
\P^\hbar_{B_k} = - \frac14 \frac{\pa \Fc_{Nek}}{\pa a_k}\,.
\ee
It is useful to rewrite the differential operators $D_i$, $\tilde{D}_i$ in terms of $\la_i$ and $m_n$:
\be
D_i = - \sum\limits_{m = 1}^{N_c} \frac{K^{(i)}(\la_m)}{K'(\la_m)} \frac{\pa}{\pa \la_m}\,,\qquad \qquad
\tilde D_i = \sum\limits_{n = 1}^{N_f} \frac{\tilde K^{(i)}(-m_n)}{\tilde K'(-m_n)} \frac{\pa}{\pa m_n}\,,
\ee
where $\tilde K = K_+ K_-$.

Then to the order $\hbar^2$ (or $\e_1^2$) one has
\begin{multline}
-\frac14 \frac{\pa \Fc_{pert}}{\pa a_i} = \sum\limits_{j \neq i}  \Bigl[ a_{ij} \l( \ln \frac{a_{ij}}{\La} -1 \r) + \frac{1}{12} \frac{\e_1^2}{a_{ij}} \Bigr] - \\
-\frac12 \sum\limits_{n} \Bigl[ (a_i+ m_n) \l( \ln \frac{a_i + m_n}{\La} - 1 \r) -\frac{1}{24} \frac{\e_1^2}{a_i + m_n} \Bigr] + o(\e_1^2) \,, \label{Ph_B}
\end{multline}
\be
\P^0_{B_i} (\P^0_{A}) = \sum\limits_{j \neq i} \la_{ij} \l( \ln \frac{\la_{ij}}{\La} -1 \r) - \frac12 \sum\limits_{n} (\la_i + m_n) \l( \ln \frac{\la_i+m_n}{\La} -1 \r)\,.\label{P0_B}
\ee
Acting by $\hat O$ on~(\ref{P0_B}) and using the identities
\bea
\tilde D_0 \ln \La &=& \frac1{N_c - \frac{N_f}{2}}\,,\\
D_2 \la_i &=& - 2 \sum\limits_{k \neq i} \frac1{\la_{ik}}\,,\\
\tilde D_2 m_n &=& - 2 \sum\limits_{m \neq n} \frac1{m_n - m_m}\,,\\
\tilde D_1 m_n &=& 1\,,\\
\sum\limits_n \sum\limits_{m \neq n} \frac{1}{m_n - m_m}&=&0\,
\eea
one gets
\begin{multline}
\hat O \P^0_{B_i} (\P^0_{A}) = \P^0_{B_i} (\P^0_{A}) - \frac{(-i\hbar)^2}{24} \l( D_2 \tilde D_0 + \tilde D_2 \tilde D_0 - \tilde D_1 \tilde D_1 \r) \P^0_{B_i} (\P^0_{A}) = \\
= p_0 + \frac{\e_1^2}{24} \Biggl[ (D_2 + \tilde D_2) \l( - \frac1{N_c - N_f/2}\r) \l( \sum\limits_{j \neq i} \la_{ij} - \frac12 \sum\limits_n (\la_i + m_n)\r)
-\tilde D_1 \l( -\frac12 \sum\limits_n \ln \frac{\la_i + m_n}{\La} \r) \Biggr] =\\
=p_0 + \frac{\e_1^2}{24} \left[ (D_2 + \tilde D_2) \l( -\la_i + \frac1{2N_c - N_f} \l( \sum\limits_n m_n + \sumlim_j \la_j \r) \r) +\frac12 \sum\limits_n \frac1{\la_i + m_n}\right]=\\
= p_0 + \frac{\e_1^2}{24} \l( \sum\limits_{j \neq i} \frac2{\la_{ij}} + \frac12 \sum\limits_n \frac1{\la_i + m_n} \r) + O(\e_1^4)\,,
\end{multline}
in agreement with~(\ref{Ph_B}). We thus check the conjecture for general $N_c$ and $N_f$ up to the order $\hbar^2 \ln \La$. Note also that we did not expand the periods in $m_f$.
\subsection{One instanton approximation}
The general statement of the conjecture
\be
\left.\frac{\pa \Fc_{Nek}}{\pa a_i}\right|_{a_i = \Oc \P^0_{A_i}(\la,\,m,\,\La)} \stackrel{?}{=} \Oc \left.\frac{\pa \Fc_{SW}}{\pa a_i}\right|_{a_i= \P^0_{A_i}(\la,\,m,\,\La)}
\ee
in the order $\hbar^2 \La^{2N_c - N_f}$ reduces to
\be
\frac{\pa \Fc_{1-inst}^{Nek''}}{\pa \la_i} + \sum\limits_{j=1}^{N_c} \frac{\pa^2 \Fc_{pert}^{Nek''}}{\pa \la_j \pa \la_i} \d \la_j \stackrel{?}{=} \sum\limits_{j=1}^{N_c} \l[ \Oc_2 \l( \d \la_j \frac{\pa^2 \Fc_{pert}^{SW}}{\pa \la_j \pa \la_i} \r) - \d \la_j \Oc_2 \frac{\pa^2 \Fc_{pert}^{SW}}{\pa \la_j \pa \la_i} \r] + \Oc_2 \frac{\pa \Fc_{1-inst}^{SW}}{\pa \la_i}\,. \label{1inst}
\ee
Here $\Oc_2 = - \frac1{24} \l(D_2 \tilde{D}_0 + \tilde{D}_2 \tilde{D}_0 - \tilde{D}_1  \tilde{D}_1\r)$, $\d \la$ is the 1-instanton correction to the classical \mbox{A-periods}~(\ref{Aperiods}) and double prime denotes the second derivative in $\hbar$ at $\hbar = 0$.
We performed a computerized check of this statement for the 1-instanton part of the Nekrasov prepotential up to the $\hbar^2$ order for lower $N_c$ and $N_f$:
\begin{itemize}
\item $N_c=2$ for $N_f=1,\,2,\,3$,
\item $N_c=3$ for $N_f=1,\,2,\,3$,
\item $N_c=4$ for $N_f=1,\,2$.
\end{itemize}
Here we present as an example relevant formulas only for the case of~$N_c=2$,~$N_f=2$.
\be
\frac{\pa \Fc_{1-inst}^{Nek''}}{\pa \la_1}=-{\frac {{\Lambda}^{2} \left( 8\,m_{{1}}m_{{2}}+3\,m_{{1}}\lambda
_{{1}}+5\,m_{{1}}\lambda_{{2}}+4\,\lambda_{{1}}\lambda_{{2}}+3\,m_{{2}
}\lambda_{{1}}+{\lambda_{{1}}}^{2}+5\,m_{{2}}\lambda_{{2}}+3\,{\lambda
_{{2}}}^{2} \right) }{ 2 \,\left( \lambda_{{1}}-\lambda_{{2}} \right) ^{5}
}}
\ee
\begin{multline}
\sum\limits_{j=1}^{2} \frac{\pa^2 \Fc_{pert}^{Nek''}}{\pa \la_j \pa \la_1} \d \la_j = \frac{{\Lambda}^{2}}{4}\, \left(  \frac1{3\,\left( \lambda_{{1}}-\lambda_{{2}} \right) ^{2}}+ \frac1{12\,\left( \lambda_{{1}}+m_{{1}} \right) ^{2}}+\frac1{12\, \left( \lambda_
{{1}}+m_{{2}} \right) ^{2}} \right) \times \\
 \times \left( {\frac {
\lambda_{{1}}+m_{{2}}}{ \left( \lambda_{{1}}-\lambda_{{2}} \right) ^{2
}}}+{\frac {\lambda_{{1}}+m_{{1}}}{ \left( \lambda_{{1}}-\lambda_{{2}}
 \right) ^{2}}}-2\,{\frac { \left( \lambda_{{1}}+m_{{1}} \right)
 \left( \lambda_{{1}}+m_{{2}} \right) }{ \left( \lambda_{{1}}-\lambda_
{{2}} \right) ^{3}}} \right)-\\
 -\frac1{12}\,\frac{{\Lambda}^{2}}{ \left( \lambda_{{1}}-\lambda_{{2}}
 \right) ^{2}} \left( {\frac {
\lambda_{{2}}+m_{{2}}}{ \left( \lambda_{{2}}-\lambda_{{1}} \right) ^{2
}}}+{\frac {\lambda_{{2}}+m_{{1}}}{ \left( \lambda_{{2}}-\lambda_{{1}}
 \right) ^{2}}}-2\,{\frac { \left( \lambda_{{2}}+m_{{1}} \right)
 \left( \lambda_{{2}}+m_{{2}} \right) }{ \left( \lambda_{{2}}-\lambda_
{{1}} \right) ^{3}}} \right)
\end{multline}
\begin{multline}
\sum\limits_{j=1}^{2} \l[ \Oc_2 \l( \d \la_j \frac{\pa^2 \Fc_{pert}^{SW}}{\pa \la_j \pa \la_i} \r) - \d \la_j \Oc_2 \frac{\pa^2 \Fc_{pert}^{SW}}{\pa \la_j \pa \la_i} \r] =\\
= \frac{{\Lambda}^{2
}}{48\,\left( \lambda_{{1}}+m_{{2}} \right) ^{2} \left( \lambda_{{1}}+m_{
{1}} \right) ^{2} \left( \lambda_{{2}}-\lambda_{{1}} \right) ^{5}} \bigl( 124\,{m_{{2}}}^{3}\lambda_{{2}}\lambda_{{1}}m_{{
1}}+51\,m_{{2}}{\lambda_{{2}}}^{2}{m_{{1}}}^{2}\lambda_{{1}}+124\,{m_{
{1}}}^{3}\lambda_{{2}}\lambda_{{1}}m_{{2}}+\\
+4\,m_{{2}}{\lambda_{{2}}}^{
3}m_{{1}}\lambda_{{1}}+336\,\lambda_{{1}}\lambda_{{2}}{m_{{1}}}^{2}{m_
{{2}}}^{2}+471\,{\lambda_{{1}}}^{2}\lambda_{{2}}{m_{{1}}}^{2}m_{{2}}+
564\,{\lambda_{{1}}}^{3}\lambda_{{2}}m_{{1}}m_{{2}}+96\,{\lambda_{{1}}
}^{2}{\lambda_{{2}}}^{2}m_{{2}}m_{{1}}+\\
+471\,{\lambda_{{1}}}^{2}\lambda
_{{2}}m_{{1}}{m_{{2}}}^{2}+18\,{m_{{1}}}^{2}{\lambda_{{2}}}^{2}{
\lambda_{{1}}}^{2}+272\,{m_{{1}}}^{3}{m_{{2}}}^{2}\lambda_{{1}}+210\,{
m_{{1}}}^{3}{\lambda_{{1}}}^{2}m_{{2}}+485\,{m_{{1}}}^{2}{\lambda_{{1}
}}^{3}m_{{2}}+\\
+272\,{m_{{1}}}^{2}{m_{{2}}}^{3}\lambda_{{1}}-{m_{{1}}}^{
3}{\lambda_{{2}}}^{2}\lambda_{{1}}+2\,{m_{{1}}}^{3}{\lambda_{{2}}}^{2}
m_{{2}}+38\,{\lambda_{{1}}}^{3}{\lambda_{{2}}}^{2}m_{{2}}+18\,{\lambda
_{{1}}}^{2}{\lambda_{{2}}}^{2}{m_{{2}}}^{2}+226\,{\lambda_{{1}}}^{4}
\lambda_{{2}}m_{{2}}+\\
+208\,{\lambda_{{1}}}^{3}\lambda_{{2}}{m_{{2}}}^{2
}+226\,{\lambda_{{1}}}^{4}\lambda_{{2}}m_{{1}}+208\,{\lambda_{{1}}}^{3
}\lambda_{{2}}{m_{{1}}}^{2}+63\,{m_{{2}}}^{3}\lambda_{{2}}{\lambda_{{1
}}}^{2}+24\,{m_{{2}}}^{2}{\lambda_{{2}}}^{2}{m_{{1}}}^{2}+64\,{m_{{2}}
}^{3}\lambda_{{2}}{m_{{1}}}^{2}+\\
+344\,m_{{2}}{\lambda_{{1}}}^{4}m_{{1}}
+485\,{m_{{2}}}^{2}{\lambda_{{1}}}^{3}m_{{1}}+648\,{m_{{2}}}^{2}{
\lambda_{{1}}}^{2}{m_{{1}}}^{2}+63\,{m_{{1}}}^{3}\lambda_{{2}}{\lambda
_{{1}}}^{2}+64\,{m_{{1}}}^{3}\lambda_{{2}}{m_{{2}}}^{2}+38\,{\lambda_{
{1}}}^{3}{\lambda_{{2}}}^{2}m_{{1}}-\\
-{m_{{2}}}^{3}{\lambda_{{2}}}^{2}
\lambda_{{1}}+2\,{m_{{2}}}^{3}{\lambda_{{2}}}^{2}m_{{1}}+210\,{m_{{2}}
}^{3}{\lambda_{{1}}}^{2}m_{{1}}+6\,{\lambda_{{1}}}^{2}{\lambda_{{2}}}^
{3}m_{{1}}+4\,\lambda_{{1}}{\lambda_{{2}}}^{3}{m_{{1}}}^{2}+6\,m_{{2}}
{\lambda_{{2}}}^{3}{\lambda_{{1}}}^{2}+m_{{2}}{\lambda_{{2}}}^{3}{m_{{
1}}}^{2}+\\
+4\,{m_{{2}}}^{2}{\lambda_{{2}}}^{3}\lambda_{{1}}+{m_{{2}}}^{2
}{\lambda_{{2}}}^{3}m_{{1}}+16\,{\lambda_{{1}}}^{4}{\lambda_{{2}}}^{2}
+106\,{m_{{2}}}^{2}{\lambda_{{1}}}^{4}+49\,{m_{{2}}}^{3}{\lambda_{{1}}
}^{3}+106\,{m_{{1}}}^{2}{\lambda_{{1}}}^{4}+\\
+49\,{m_{{1}}}^{3}{\lambda_
{{1}}}^{3}+84\,{\lambda_{{1}}}^{5}\lambda_{{2}}+66\,m_{{2}}{\lambda_{{
1}}}^{5}+66\,m_{{1}}{\lambda_{{1}}}^{5}+4\,{\lambda_{{1}}}^{3}{\lambda
_{{2}}}^{3}+{m_{{1}}}^{3}{\lambda_{{2}}}^{3}+\\
+{m_{{2}}}^{3}{\lambda_{{2
}}}^{3}+8\,{\lambda_{{1}}}^{6}+51\,\lambda_{{1}}{\lambda_{{2}}}^{2}{m_
{{2}}}^{2}m_{{1}}+112\,{m_{{1}}}^{3}{m_{{2}}}^{3} \bigr)
\end{multline}
\begin{multline}
\Oc_2 \frac{\pa \Fc_{1-inst}^{SW}}{\pa \la_1} ={\frac {{\Lambda}^{2} \left( 4\,m_{{1}}\lambda_{{2}}+2\,m_{{1}}\lambda_{{1}}+6
\,m_{{1}}m_{{2}}+3\,{\lambda_{{2}}}^{2}+{\lambda_{{1}}}^{2}+2\,\lambda
_{{1}}\lambda_{{2}}+4\,m_{{2}}\lambda_{{2}}+2\,m_{{2}}\lambda_{{1}}
 \right) }{3\, \left( \lambda_{{2}}-\lambda_{{1}} \right) ^{
5}}}
\end{multline}
Here double prime denotes the second derivative in $\e_1$ at $\e_1=0$. Summing up all the contributions one can see that the left hand side of Eq.~(\ref{1inst}) indeed equals the right hand side in this case.
\begin{figure}[t]
\begin{center}
  \includegraphics[width=14cm]{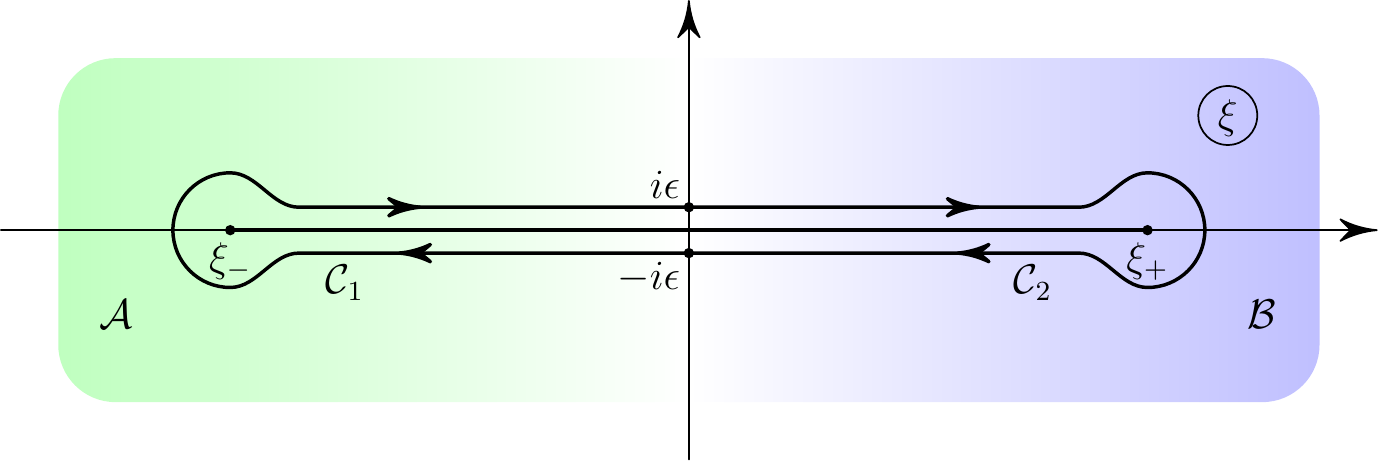}\\
  \caption{The B-contour $\Cc$ consisting of two stretches $\Cc_1$ and $\Cc_2$ encircles the cut shown in the complex $\xi$ plane.}\label{B-cont}
\end{center}
\end{figure}

\section{Exact one-loop part in the $SU(2)$ case}
\label{exactcheck}
In this section we reproduce the one-loop part of the Nekrasov prepotential exactly to all orders in $\hbar$. The periods in this case are expressed through the Harish-Chandra functions as was noted in~\cite{Marshakov:2010fx}. Additionally we show directly that moving the hypermultiplets from $K_+$ to $K_-$ polynomials in the Baxter equation~(\ref{equation}) does not affect quantum periods in the perturbative limit.
\subsection{Pure gauge theory}
Let us consider first the pure $SU(2)$ case. The Baxter equation in this case reads
\be
(-i\hbar \pa_\xi)^2 \psi + \la^2 \psi + \frac{\bar{\La}}{2} \l( e^\xi + e^{-\xi} \r) \psi = 0\,. \label{su2}
\ee
where $\xi = ix$.
This equation is just the Schr\"odinger equation describing the motion of a particle in the potential well $V(\xi) = \bar \La \ch \xi$ with energy $E=-\la^2$ ($\la$ is assumed to be pure imaginary with $\Im \la >0$). The classically allowed motion is confined between the turning points $\bar{\La} \ch \xi_{\pm} = -\la^2$. The B-periods featuring in the Bohr-Sommerfeld quantization conditions are calculated along the contour $\Cc$ encircling the turning points $\xi_{\pm}$ (see Figure~\ref{B-cont}).

The value of the B-period is equal to the phase acquired by the particle moving around one period of oscillations. Let us consider the particle starting from $\xi=0$ in the leftwards direction, reflecting from the potential wall on the left and returning to $\xi=0$ point. The phase $\phi_1$ acquired in such a process is given by the ratio of the coefficients in front of the incoming and outgoing waves in the particle's wavefunction $\psi$:
\bea
\psi(\xi \to 0) &=& c_+ e^{\frac{\la}{\hbar}\xi} + c_- e^{-\frac{\la}{\hbar}\xi} \,,\\
\phi_1 &=& \frac{i}{\hbar} \int\limits_{\Cc_1} P(\x) \, d\x  = \ln \frac{c_+}{c_-}\,.
\eea
The same phase shift $\phi_2$ is acquired while traveling the other half of the period
\be
\phi_2 = \frac{i}{\hbar} \int\limits_{\Cc_2} P(\x) \, d\x  = \ln \frac{c_+}{c_-}\,.
\ee
Should the energy $E_n =-\la^2$ be a true energy level of the system, the wavefunction will be single-valued and thus the phase shift will be $2 \p n$ with $n \in \mathbb{Z}$. For general $\la$ this is not the case, the wave function is multivalued and the phase shift (or monodromy) gives the B-periods:
\be
\P^\hbar_B = \int\limits_\Cc P(x) \, dx = -i \int\limits_\Cc P(\x)\, d\x = -i \int\limits_{\Cc_1} P(\x) \, d\x -i \int\limits_{\Cc_2} P(\x)\, d\x = -2\hbar \ln \frac{c_+}{c_-} \label{su2per}
\ee

In the perturbative limit $\bar{\La} \to 0$ the turning points $\xi_{\pm}$ move to $\pm \infty$ and thus the particle moves freely most of the time, but encounters the exponential potential walls on the left and right.
To determine the wavefunction in this case one must solve the equation~(\ref{su2}) separately in the regions $\Ac$ and $\Bc$ close to the turning points $\xi_{\pm}$ respectively.
\be
\begin{array}{cc}
\mbox{On the left:} & \mbox{On the right:}\\
\x \simeq \x_- \simeq \ln \frac{\bar{\La}}{2}\,, & \x \simeq \x_+ \simeq -\ln \frac{\bar{\La}}{2}\,,\\
\tilde{\x} = \x - \ln \frac{\bar{\La}}{2}\,, & \tilde{\x} = \x + \ln \frac{\bar{\La}}{2}\,,\\
(-i\hbar \pa)^2 \psi_\Ac + \la^2 \psi_\Ac + e^{-\tilde{\xi}} \psi_\Ac = 0\,, & (-i\hbar \pa)^2 \psi_\Bc + \la^2 \psi_\Bc + e^{\tilde{\xi}} \psi_\Bc = 0\,,\\
\psi_\Ac (\tilde{\xi} \to -\infty) = 0\,, & \psi_\Bc (\tilde{\xi} \to \infty) = 0\,.
\end{array}\notag
\ee
The solutions are given by
\be
\begin{array}{cc}
\mbox{On the left:} & \mbox{On the right:}\\
\psi_\Ac = f_0(\la;-\x) \stackrel{\text{def}}{=} K_{\frac{2\la}{\hbar}} \l( \sqrt{\frac{\bar{\La}}{2}} \frac{2e^{-\x/2}}{\hbar}\r)\,, & \psi_\Bc = f_0(\la;\xi)\,.
\end{array} \notag
\ee
The asymptotics for $\xi \to 0$ are
\be
f_0(\la;\xi \to 0) = \frac1{\G \l( 1-\frac{2\la}{\hbar} \r)} \l( \frac{\bLa}{2 \hbar^2}\r)^{-\frac{\la}{\hbar}} e^{\frac{\la}{\hbar} \xi} - \frac1{\G \l( 1+\frac{2\la}{\hbar} \r)} \l( \frac{\bLa}{2 \hbar^2}\r)^{\frac{\la}{\hbar}} e^{-\frac{\la}{\hbar} \xi}\,,
\ee
and according to Eq.~(\ref{su2per}) the B-periods are given by
\be
\P_B^\hbar = - 2\hbar \ln \l[ - \l( \frac{\bLa}{2\hbar^2}\r)^{-\frac{2\la}{\hbar}} \frac{\G \l(1+ \frac{2\la}{\hbar} \r)}{\G \l(1 - \frac{2\la}{\hbar} \r)} \r]\,.
\ee
Up to the terms linear in $\la$ this coincides with the (relative) perturbative period determined by the Nekrasov function Eq.~(\ref{Nekpert}):
\be
\P_B^\hbar = \frac{1}{2}  \l[ \frac{\pa \Fc_{Nek}}{\pa a_2} - \frac{\pa \Fc_{Nek}}{\pa a_1} \r] \Biggr|_{\begin{array}{l} a_1=-\la \\ a_2=\la \end{array}}\,.
\ee
\subsection{One hypermultiplet}
Analysis of the $SU(2)$ gauge theory with fundamental hypermultiplets runs parallel with the previous subsection. The equation now is
\be
(-i\hbar \pa_\xi)^2 \psi + \la^2 \psi + \frac{\bar{\La}}{2} \l( e^\xi \l(\hbar \pa_\x + m + \frac{\hbar}{2} \r) + e^{-\xi} \r) \psi = 0\,. \label{su2+1}
\ee
We have put the hypermultiplet in front of the first exponent. We will see that the other choice gives the same period. Again in the limit $\bLa \to 0$ one must solve the equation separately in two regions $\Ac$ and $\Bc$
\be
\begin{array}{cc}
\mbox{On the left:} & \mbox{On the right:}\\
\x \simeq \x_- \simeq \ln \frac{\bar{\La}}{2}\,, & \x \simeq \x_+ \simeq -\ln \frac{\bar{\La}}{2}\,,\\
\tilde{\x} = \x - \ln \frac{\bar{\La}}{2}\,, & \tilde{\x} = \x + \ln \frac{\bar{\La}}{2}\,,\\
(-i\hbar \pa)^2 \psi_\Ac + \la^2 \psi_\Ac + e^{-\tilde{\xi}} \psi_\Ac = 0\,, & (-i\hbar \pa)^2 \psi_\Bc + \la^2 \psi_\Bc + e^{\tilde{\xi}} \l(\hbar \pa + m + \frac{\hbar}{2} \r) \psi_\Bc = 0\,,\\
\psi_\Ac (\tilde{\xi} \to -\infty) = 0\,, & \psi_\Bc (\tilde{\xi} \to \infty) = 0\,.
\end{array}
\ee
The solutions are
\be
\psi_\Ac = f_0(\la;-\x)\,, \label{1sol1}\\
\ee
\begin{multline}
\psi_\Bc = f_1(\la, m;\xi) \stackrel{\text{def}}{=} \l(\frac{\bLa}{2\hbar}\r)^{\frac{\la}{\hbar}} e^{\frac{\la}{\hbar} \x} \Phi \l( \frac{\la+m}{\hbar} + \frac12, 1+ \frac{2\la}{\hbar} ; {\frac{\bLa}{2\hbar}} e^\x \r) - \\
- \l(\frac{\bLa}{2\hbar}\r)^{-\frac{\la}{\hbar}} e^{-\frac{\la}{\hbar} \x} \frac{\G \l(1+ \frac{2\la}{\hbar} \r)   \G \l(\frac12+ \frac{m-\la}{\hbar} \r)}{\G \l(1 - \frac{2\la}{\hbar} \r)  \G \l(\frac12+ \frac{m+\la}{\hbar} \r)} \Phi \l( \frac{m-\la}{\hbar} + \frac12, 1 - \frac{2\la}{\hbar} ; {\frac{\bLa}{2\hbar}} e^\x \r)\,, \label{1sol2}
\end{multline}
where $\Phi(a,c;z)$ is degenerate hypergeometric function. The situation in the present case is clearly not left-right symmetric. Thus the phases acquired on two halves of the period are not equal. If we introduce $c_{\pm}^L$, $c_{\pm}^R$ as
\bea
\psi_\Ac(\x \to 0) &=& c_+^L e^{\frac{\la}{\hbar} \x} + c_-^L e^{-\frac{\la}{\hbar} \x}\,,\\
\psi_\Bc(\x \to 0) &=& c_+^R e^{\frac{\la}{\hbar} \x} + c_-^R e^{-\frac{\la}{\hbar} \x}\,,
\eea
then analogously to Eq.~(\ref{su2per}) the period will be given by
\be
\P_B^\hbar = -\hbar \ln \l[ \frac{c^L_+}{c^L_-} \frac{c^R_-}{c^R_+} \r]\,.   \label{asym}
\ee
Using Eq.~(\ref{1sol1}),~(\ref{1sol2}) one gets
\be
\P_B^\hbar = -\hbar \ln \l[- \l( \frac{\bLa}{2} \r)^{-\frac{4\la}{\hbar}} \hbar^{\frac{6\la}{\hbar}} \l( \frac{\G \l(1+ \frac{2\la}{\hbar} \r)}{\G \l(1 - \frac{2\la}{\hbar} \r)}\r)^2 \cdot \, \frac{\G \l(\frac12+ \frac{m-\la}{\hbar} \r)}{  \G \l(\frac12+ \frac{m+\la}{\hbar} \r)} \r]\,.
\ee
This again coincides with the (relative) periods from Eq.~(\ref{Nekpert}). Now let us try to put the hypermultiplet in front of the second exponent in Eq.~(\ref{su2+1}):
\be
(-i\hbar \pa_\xi)^2 \psi + \la^2 \psi + \frac{\bar{\La}}{2} \l( e^\xi  + e^{-\xi} \l(\hbar \pa_\x + m - \frac{\hbar}{2} \r) \r) \psi = 0\,.
\ee
The solutions in regions $\Ac$ and $\Bc$ are related to the solutions with different placement of the multiplet:
\bea
\psi'_\Ac (\xi) &=& \exp \l( - \frac{\bLa e^{-\x}}{2\hbar}\r) f_1(\la,m;-\xi) \,,\\
\psi'_\Bc (\xi) &=& f_0(\la;\xi)\,.
\eea
Clearly the B-periods remain the same in this case.
\subsection{Two hypermultiplets}
There are four ways to place the hypermultiplets in the Baxter equation: both can stick either to the first or the second exponents. We first examine the ``symmetric'' case with one hypermultiplet on the right and one on the left:
\be
(-i\hbar \pa_\xi)^2 \psi + \la^2 \psi + \frac{\bar{\La}}{2} \l( e^\xi \l(\hbar \pa_\x + m_1 + \frac{\hbar}{2} \r) + e^{-\xi} \l(\hbar \pa_\x + m_2 - \frac{\hbar}{2} \r)\r) \psi = 0\,. \label{su2+2a}
\ee
The solutions in the regions $\Ac$ and $\Bc$ are related to the solution~(\ref{1sol2}) from the previous subsection
\bea
\psi_\Ac (\xi) &=& \exp \l( - \frac{\bLa e^{-\x}}{2\hbar}\r) f_1(\la,m_2;-\xi)\,, \\
\psi_\Bc (\xi) &=& f_1(\la,m_1;\xi)\,.
\eea
According to Eq.~(\ref{asym}) The period is
\be
\P_B^\hbar = -\hbar \ln \l[ \l( \frac{\bLa}{2\hbar} \r)^{-\frac{4\la}{\hbar}} \l( \frac{\G \l(1+ \frac{2\la}{\hbar} \r)}{\G \l(1 - \frac{2\la}{\hbar} \r)}\r)^2 \cdot \, \frac{\G \l(\frac12+ \frac{m_1-\la}{\hbar} \r)}{  \G \l(\frac12+ \frac{m_1+\la}{\hbar} \r)} \cdot \frac{\G \l(\frac12+ \frac{m_2-\la}{\hbar} \r)}{  \G \l(\frac12+ \frac{m_2+\la}{\hbar} \r)}  \r]\,,
\ee
which is in agreement with Eq.~(\ref{Nekpert}). The answer is clearly symmetric in the masses of the multiplets. Therefore exchanging them in Eq.~(\ref{su2+2a}) does not affect the period.

Now we try out the ``asymmetric'' form of the Baxter equation:
\be
(-i\hbar \pa_\xi)^2 \psi + \la^2 \psi + \frac{\bar{\La}}{2} \l( e^\xi \l(\hbar \pa_\x + m_1 + \frac{\hbar}{2} \r) \l(\hbar \pa_\x + m_2 + \frac{\hbar}{2} \r) + e^{-\xi} \r) \psi = 0\,. \label{su2+2b}
\ee
The solution in the $\Ac$ region is the same as in the pure gauge theory: $\psi_\Ac = f_0(\la;-\x)$; in the $\Bc$ region Eq.~(\ref{su2+2b}) becomes
\be
(-i\hbar \pa)^2 \psi_\Bc + \la^2 \psi_\Bc + e^{\tilde{\x}} \l(\hbar \pa + m_1 + \frac{\hbar}{2} \r) \l(\hbar \pa + m_2 + \frac{\hbar}{2} \r) \psi_\Bc = 0\,,
\ee
with $\tilde{\x} = \x + \ln \bLa/2$. The solution is chosen so as to be regular at the point $e^{\tilde{\x}}=1$:
\begin{multline}
\psi_\Bc = f_2(m_1,m_2,\la;\xi) \stackrel{\text{def}}{=} \frac{\G \l( 1 - \frac{2\la}{\hbar}\r) }{\G \l( \frac{m_1 - \la}{\hbar} + \frac{1}{2}\r) \G \l( \frac{m_2 -\la}{\hbar} +\frac12\r)} \cdot \\
\cdot\l( \frac{\bLa}{2}\r)^{\frac{\la}{\hbar}} e^{\frac{\la}{\hbar} \x} {}_2F_1 \l(\frac{\la+m_1}{\hbar} +\frac12,\frac{\la+m_2}{\hbar} +\frac12;1+\frac{2\la}{\hbar}; \frac{\bLa}{2} e^\x \r) -\\
- \frac{\G \l( 1 + \frac{2\la}{\hbar}\r) }{\G \l( \frac{m_1 + \la}{\hbar} +\frac12 \r) \G \l( \frac{m_2 + \la}{\hbar} +\frac12\r)} \l( \frac{\bLa}{2} \r)^{-\frac{\la}{\hbar}} e^{-\frac{\la}{\hbar} \x} {}_2F_1 \l(\frac{m_1-\la}{\hbar} +\frac12,\frac{m_2-\la}{\hbar} +\frac12;1-\frac{2\la}{\hbar}; \frac{\bLa}{2} e^\x \r)\,.
\end{multline}
Here $_2F_1(a,b;c;z)$ is the hypergeometric function. The period determined by the asymptotics of the solutions $\psi_\Ac$, $\psi_\Bc$ coincides with the one obtained in the ``symmetric'' form of the Baxter equation. One also obtains the same result with yet another placement of the multiplets:
\be
(-i\hbar \pa_\xi)^2 \psi + \la^2 \psi + \frac{\bar{\La}}{2} \l( e^\xi + e^{-\xi}  \l(\hbar \pa_\x + m_1 - \frac{\hbar}{2} \r) \l(\hbar \pa_\x + m_2 - \frac{\hbar}{2} \r)\r) \psi = 0\,.
\ee
The solutions are
\bea
\psi_\Ac (\x) &=& f_2(\la, -m_1, -m_2; - \x)\,,\\
\psi_\Bc (\x) &=& f_0(\la,\x)\,.
\eea
Thus the periods are the same in this case. Summing up, we see that shuffling the hypermultiplets between the two terms of the Baxter equation indeed does not affect the periods.
\subsection{Three hypermultiplets}
Two of the three hypermultiplets must be placed in front of one of the exponents. Otherwise the resulting equation will be higher than second order and therefore not tractable. Following the same steps as in the previous subsections one gets the solutions and the period for this case:
\bea
\psi_\Ac (\x) &=& \exp \l( - \frac{\bLa e^{-\x}}{2\hbar}\r) f_1(\la, m_3; - \x)\,,\\
\psi_\Bc (\x) &=& f_2(\la, m_1, m_2; \x)\,.
\eea
\be
\P_B^\hbar = -\hbar \ln \l[ \l( \frac{\bLa}{2\sqrt{\hbar}} \r)^{-\frac{4\la}{\hbar}} \l( \frac{\G \l(1+ \frac{2\la}{\hbar} \r)}{\G \l(1 - \frac{2\la}{\hbar} \r)}\r)^2 \cdot \, \prodlim_{f=1}^3 \frac{\G \l(\frac12+ \frac{m_f-\la}{\hbar} \r)}{  \G \l(\frac12+ \frac{m_f+\la}{\hbar} \r)} \r]\,.
\ee
\subsection{Four hypermultiplets}
For the case of four hypermultiplets instead of $\bLa$ one has $q_{\text{UV}} = \exp(2\p i \t_{\text{UV}})$ with $\t_{\text{UV}}$ being the microscopic coupling, well defined in this situation. Otherwise this case is in complete analogy with the previous ones and the periods are found to be
\be
\P_B^\hbar = -\hbar \ln \l[ \l( \frac{q_{\text{UV}}}{2} \r)^{-\frac{4\la}{\hbar}} \l( \frac{\G \l(1+ \frac{2\la}{\hbar} \r)}{\G \l(1 - \frac{2\la}{\hbar} \r)}\r)^2 \cdot \, \prodlim_{f=1}^4 \frac{\G \l(\frac12+ \frac{m_f-\la}{\hbar} \r)}{  \G \l(\frac12+ \frac{m_f+\la}{\hbar} \r)} \r]\,.
\ee
\section{Conclusions}
\label{conclusions}
We have presented evidence that Nekrasov prepotential for $SU(N)$ theory with hypermultiplets in fundamental representation arises when one considers the quantization of the corresponding Seiberg-Witten curve. We have checked this conjecture:
\begin{itemize}
\item For perturbative part of the prepotential for general $N_c$ and $N_f$ to the order $\hbar^3$.
\item For 1-instanton part for $N_c=1,\,2$, $N_f=1,\,2,\,3$ and $N_c=4$, $N_f=1,\,2$ to the order $\hbar^3$.
\item For perturbative part of the prepotential for $N_c=2$ and $N_f=0,\,1,2,4$ to all orders in $\hbar$.
\end{itemize}
What is important, these checks were exact in masses of the hypermultiplets. Thus we have also checked that the Baxter equation for the XXX spin chain and Gaudin magnet give the same quantum deformed periods in the listed cases. This statement requires further investigation because for the Gaudin system only lowest correction in~$\hbar$ for the case of $SU(2)$ was obtained in~\cite{Maruyoshi:2010iu}.
\section*{Acknowledgements}
Author is grateful to A. Morozov and A. Mironov for stimulating discussions and advice. This work was supported by grants RFBR 10-02-00499, RFBR-Royal Society 11-01-92612, by Ministry of Education and Science of the Russian Federation under contract 02.740.11.5194 and by Dynasty foundation.

\textbf{Note.} After finishing this paper we grew aware of the recent preprint~\cite{Fucito:2011pn} in which related subjects are discussed.

\end{document}